\documentclass[wsdraft]{ws-ijmpd}

\usepackage[T1]{fontenc}
\usepackage[full]{textcomp}  
\usepackage{stix}

\usepackage{cite}
\usepackage{amsmath,amssymb,graphicx}
\usepackage{babel}
\usepackage{graphicx,color}
\usepackage{braket}
\usepackage{mathtools}
\usepackage{slashed}
\usepackage{empheq}
\usepackage{tikz}
\usepackage{multirow}

\begin{document}

\markboth{J.J. Ramos and J.E.G Silva}
{Braneworld cosmology in $f(\mathbb{Q})$ gravity}

%
\catchline{}{}{}{}{}
%

\title{Braneworld cosmology in $f(\mathbb{Q})$ gravity}

\author{ J.J. Ramos}

\address{{Universidade Federal do Cear\'{a} (UFC), Departamento de F\'{i}sica, Campus do Pici, Fortaleza-CE, 60455-760, Brazil.}\\
joaojardel@fisica.ufc.br}

\author{J.E.G Silva}

\address{{Universidade Federal do Cear\'{a} (UFC), Departamento de F\'{i}sica, Campus do Pici, Fortaleza-CE, 60455-760, Brazil.}\\
euclides@fisica.ufc.br}

\maketitle


\begin{abstract}
 This work investigates the cosmology of a thick brane within the context of $f(\mathbb{Q})$ gravity, an extension of symmetric teleparallelism. Using a five-dimensional Friedmann-Lemaître-Robertson-Walker metric, we solve the field equations to obtain dynamic solutions for the scale factor. We demonstrate that the effective cosmological constant on the brane naturally emerges as a function of the extra dimension $\Lambda(y)$, being both generated and confined by the curved geometry of the bulk. We analyze two distinct regimes: Randall-Sundrum-type thin brane and thick brane through the Sine-Gordon model. Our model reproduces accelerated expansion solutions without requiring the introduction of a fundamental cosmological constant on the brane, showing that the cosmic acceleration emerges as a consequence of the brane's embedding and the gravitational dynamics in the bulk      . The variation of the $c_i$ parameters in symmetric teleparallelism enables different cosmological scenarios, including de Sitter-type expansion, contraction, and oscillatory solutions. The results indicate that the brane's position in the bulk determines its cosmology, providing a geometric explanation for the smallness of the observed cosmological constant.
\end{abstract}


  \section{Introduction}
 
Although modern cosmology achieved many successes in the second half of the 20th century, it still faces significant challenges in describing the universe. These challenges include the cosmological constant \cite{SupernovaSearchTeam:1998fmf}, dark energy, and dark matter \cite{Peebles:2002gy,Boehm:2000gq}, inflation \cite{Guth:1980zm}, and even determining the universe's own geometry \cite{DiValentino:2019qzk}. In recent years, cosmologists have focused on understanding the tensions and cosmological anomalies that have arisen in the most recent observational data \cite{Abdalla:2022yfr,Perivolaropoulos:2021jda}.

Over the years, we have had many suggestions on how to address these problems. Proposed solutions range from modifying Einstein's general theory of relativity to the inclusion of extra dimensions in the universe.

In the scientific literature, there are several suggestions on how we can modify Einstein's theory of gravity. Among these alternatives, we have theories with additional scalar and vector degrees of freedom \cite{Capozziello:2011et}, theories based on massive gravitons \cite{Hinterbichler:2011tt}, and theories with extra dimensions, inspired by string theories, such as brane-world models \cite{Maartens:2010ar}. There are also alternative geometric constructions to Riemannian geometry, on which general relativity was built. In this latter case, we can identify theories based on Einstein-Cartan geometry \cite{Hehl:1976kj}, metric-affine (or Palatini) theories of different kinds \cite{Hehl:1994ue}, including $f(R)$ models \cite{DeFelice:2010aj}, and also theories whose dynamics is entirely based on torsion or in non-metricity \cite{Arcos:2004zh,BeltranJimenez:2017tkd}.

Theories based on torsion or non-metricity belong to a class of theories known as teleparallel theories. In such theories, spacetime is conceived with geometry that does not exhibit curvature, meaning that the source of gravity must be replaced by another property of space. For example, we have the teleparallel equivalent of general relativity (TEGR), where the source of gravitational interaction is the torsion of spacetime. Unlike general relativity, where torsion is null and spacetime curvature generates gravity, in TEGR, curvature is null and spacetime torsion generates gravitational dynamics. In this theory, the dynamic variable is the tetrad fields (vierbeins), in contrast to general relativity, where it was the metric tensor. The connection built on the condition of zero curvature and in terms of tetrads is known as the Weitzenböck connection \cite{Aldrovandi:2013wha}. Originally proposed by Einstein to attempt to geometricize the electromagnetic field, TEGR ended up becoming a gauge theory for gravity based on the Poincaré symmetry \cite{Baez:2012bn,Hohmann:2017duq}. Similarly to $f(R)$ theories, modified theories of TEGR have also been explored, such as theories where the Lagrangian density of the theory is rewritten in terms of a scalar $T$, giving rise to $f(T)$ theories \cite{Ferraro:2006jd,Cai:2015emx}.

Alongside TEGR, we have Symmetric Teleparallel Equivalent to General Relativity (STEGR). In this theory, both curvature and torsion are null, and the source of gravitational interaction lies in spacetime non-metricity, namely, in the variation of the length of vectors that are transported parallelly along a closed trajectory. Proposed over two decades ago in \cite{Nester:1998mp}, it did not attract much attention from physicists until recently when a veritable flood of articles filled scientific journals in the fields of gravitation and cosmology. In this theory, the dynamics of the gravitational field are described in terms of the non-metricity tensor $Q_{\rho\mu\nu}=\nabla_{\rho}g_{\mu\nu}$; once again, the theory's dynamic variable is the metric tensor, and therefore the theory is symmetric \cite{BeltranJimenez:2018vdo}.

Extensions to STERG can also be made by introducing a modified theory of gravity called $f(\mathbb{Q})$, where the Lagrangian density is written in terms of a function of the non-metricity scalar $\mathbb{Q}$ \cite{BeltranJimenez:2017tkd,Xu:2019sbp}. This extension has been applied in various gravitational contexts, such as early and late cosmology \cite{BeltranJimenez:2019tme,Lymperis:2022oyo,Lu:2019hra,Atayde:2021pgb}, bouncing cosmology \cite{Bajardi:2020fxh,Agrawal:2022vdg,Gadbail:2023loj}, quantum cosmology \cite{Bajardi:2023vcc,Dimakis:2021gby}, black holes \cite{Lin:2021uqa,DAmbrosio:2021zpm,Junior:2023qaq}, wormholes \cite{Mustafa:2021ykn,Mishra:2023bfe}, and even in the scenario of thick branes \cite{Silva:2022pfd,Fu:2021rgu,Belchior:2023xgn}. For an excellent pedagogical introduction to symmetric teleparallelism and its extensions, see \cite{Heisenberg:2023lru}.

As mentioned in the first paragraph of this introduction, one possible approach to address some of the previously mentioned issues in modern cosmology is to consider the existence of extra dimensions in the universe. Among the theories involving extra dimensions, we have brane-world models, which propose the existence of a four-dimensional hypersurface called a brane, where matter fields are confined, embedded in a higher-dimensional space known as the bulk, where only gravity can propagate. The brane-world model was originally proposed as a geometric solution to the hierarchy problem of fundamental forces by Lisa Randall and Raman Sundrum \cite{Randall:1999ee}. In the Randall-Sundrum (RS) models, the brane is considered infinitely thin, and the bulk is chosen as an $AdS_5$ spacetime. Subsequently, thick brane-world models were proposed as domain walls with one or more scalar fields as sources \cite{DeWolfe:1999cp}.
 
As the brane models were being developed, the cosmological implications of these theories began to be explored. We can study the cosmology of brane worlds through two formalisms: the "brane-based" formalism and the "bulk-based" formalism, both of which are entirely equivalent. In the "brane-based" formalism, the bulk geometry is projected onto the brane using the Gauss-Codazzi equations \cite{Shiromizu:1999wj, Binetruy:1999ut} and the Israel junction conditions \cite{Israel:1966rt}. Through the Gauss-Codazzi equations, the intrinsic curvature of the bulk is related to the intrinsic and extrinsic curvatures of the brane, and the induced Einstein equations on the brane are obtained. In the "bulk-based" formalism, we have a dynamic brane in a static bulk. The brane metric must be a dynamic metric with a high degree of symmetry, such as the FLRW metric, and the Einstein equation can then be solved for the bulk geometry, thus obtaining the equations of motion. The cosmology of thick branes can be satisfactorily analyzed through the bulk-based formalism.

In recent works, the cosmology of five-dimensional scenarios has been analyzed in the context of thick branes following a bulk-based formalism, where the four-dimensional part contains an FLRW universe geometry \cite{Ahmed:2013lea, Bernardini:2014vba, daSilva:2016ntp}. The general methodology of these works begins with a five-dimensional scalar field that is a solution for a brane universe where the warp factor, given as a function of conformal time and the extra dimension, is separable. Dynamic solutions for the four-dimensional brane can then be obtained.

In this work, we present a continuation of previous studies by exploring the cosmology of a thick brane within a modified theory of symmetric teleparallel gravity, namely $f(\mathbb{Q})$ gravity. We provide dynamic solutions for the case where the scalar field generating the brane depends on the extra dimension but not on time. Our results reveal solutions that exhibit accelerated expansion without requiring the introduction of a cosmological constant on the brane. Consequently, the observed cosmic acceleration can be interpreted as a geometric effect arising from the propagation of gravity into the extra dimension. We demonstrate that the effective cosmological constant on the brane can be expressed in terms of the bulk cosmological constant and extra-dimensional contributions. Subsequently, we examine the thin brane limit, deriving a fine-tuning condition that relates the brane's cosmological constant, the bulk cosmological constant, and the $c_i$ parameters of symmetric teleparallelism. Finally, we investigate the thick brane regime using Bogomolnyi-Prasad-Sommerfield (BPS) solutions for the extra-dimensional terms. Through the specific choice of the Sine-Gordon model, we establish a direct correspondence between the effective cosmological constant and the parameters of this theoretical framework. We now observe that by treating the effective cosmological constant as a function of the extra dimension, the position of the brane within the bulk determines its cosmology, since the value of the cosmological constant on the brane depends on its specific location in the bulk.

This work is organized as follows. In Section 2, we introduce the concept of symmetric teleparallelism, define the main relationships and formulas describing this geometric framework, and finally present $f(\mathbb{Q})$ gravity. In Section 3, we study the cosmology of a thick brane generated by a real scalar field in $f(\mathbb{Q})$ gravity. Using the five-dimensional Friedmann-Lemaître-Robertson-Walker metric, we solve the field equations of $f(\mathbb{Q})$ gravity and obtain the scale factor for the case where the scalar field depends only on the extra dimension. For this purpose, we propose two scenarios for the warp factor and the scalar field: the Randall-Sundrum-type thin brane regime and the thick brane scenario through the Sine-Gordon model. Additionally, we discuss how the curved bulk space influences the effective cosmological constant on the brane, and we analyze further modifications in the model arising from the parameter choices in symmetric teleparallelism. We conclude in Section 4 with a discussion of our results and future perspectives

\section{Symmetric Teleparalelism and $f(\mathbb{Q})$ gravity}

In the theory of general relativity, gravitational interaction manifests due to the curvature of spacetime, which is encoded by the Ricci tensor $R_{MN}$.
However, there are alternative formulations of relativity that yield equations equivalent to Einstein's field equations. One of these formulations is symmetric teleparallelism equivalent to general relativity (STERG)\cite{Nester:1998mp}. In this theory, the source of gravitational interaction lies in the non-metricity of space, meaning the variation of the length of vectors that are transported parallelly along a closed trajectory. This variation is described by the non-metricity tensor,  defined by \cite{BeltranJimenez:2018vdo}
\begin{equation}
\label{Tensor_de_não_metricidade}
Q_{KMN} \equiv \nabla_{K} g_{MN}.
\end{equation}
However, in  general relativity, we also learn that this theory is compatible with the metric, meaning the derivative of the metric tensor must be zero. Although, this is only true if the connection is the Levi-Civita connection. Therefore,
\begin{equation}
  \nabla_{M} g_{NP} = \partial_{M}g_{NP} - \Gamma_{MN}^{O}  g_{OP}-\Gamma_{MN}^{O} g_{NO}=0,    
\end{equation}
if, 
\begin{equation}
\label{levi-civita}
 \Gamma_{MN}^{O}=\left\{ \stackrel{O}{_{ M \ N}}\right\}= \frac{1}{2} g^{O K} \left( \partial_{M} g_{K N} + \partial_{N} g_{K M} - \partial_{K} g_{M N} \right).
\end{equation}

An interesting result is that the connection can be decomposed as \cite{BeltranJimenez:2018vdo}:
\begin{equation}
\Gamma^{O}_{MN}=\left\{ \stackrel{O}{_{ M \ N}}\right\}+ L^{O}\ _{MN} + K^{O}\ _{MN},
\end{equation}  
where $\left\{ \stackrel{O}{{ _M \ _N}} \right\}$ is the Levi-Civita connection defined by (\ref{levi-civita}), $L^{O}\ _{MN}$ is called the deformation tensor, and $K^{O}\ _{MN}$ is the cotorsion tensor, defined respectively as \cite{BeltranJimenez:2018vdo}
\begin{align}
\label{def_deformação}
L^{O}\ _{MN} \equiv \frac{1}{2}Q^{O}\ _{MN} - Q\stackrel{O}{ _{(M \ N)}}\\
\label{def_cotorção}
K^{O}\ _{MN} \equiv \frac{1}{2}T^{O}\ _{MN} + T\stackrel{O}{ _{(M \ N)}}.
\end{align}
The deformation tensor (\ref{def_deformação}) is written in terms of the non-metricity tensor (\ref{Tensor_de_não_metricidade}), 
and the tensor (\ref{def_cotorção}) is defined in terms of the torsion tensor:
\begin{equation}
T^{O}\ _{NP}\equiv 2 \Gamma^{O}_{[NP]}.
\end{equation}

 From the non-metricity tensor, we define the trace vectors as \cite{BeltranJimenez:2018vdo}:
For convenience, we introduce the superpotential tensor \cite{BeltranJimenez:2018vdo, BeltranJimenez:2019tme} as follows:
\begin{align}
\label{superpotencial}
P^{K\ }_{MN}=c_1Q^{K}\ _{MN} + c_2  Q\stackrel{K}{ _{(M \ N)}} +c_3Q^{K}g_{MN} +c_4 \delta^{K} _{(M}\stackrel{_{-}}{Q}_{N)} \\ \nonumber +\dfrac{c_5}{2}\left(\stackrel{_{-}}{Q^{K}}g_{MN} +\delta^{K} _{(M} Q_{N)}\right), 
\end{align}

where $c_i$ are constants. We can define a quadratic invariant called the generalized non-metricity scalar by contracting $P^{K}\ _{MN}$ and $Q_{K}\ ^{MN}$ as follows:
\begin{equation}
\label{escalardenãometricidade}
\mathbb{Q} =Q_{T}\ ^{MN} P^{T}\ _{MN}.
\end{equation}

Now, considering the notation developed above, the Ricci scalar for the connection $\Gamma^{O}_{MN}$ can be written as:
\begin{equation}
R=\mathcal{R}+\mathbb{Q}+D_{\kappa}(Q^{\kappa}-\stackrel{_{-}}{Q^{\kappa}}),
\end{equation}
where $\mathcal{R}$ is the Ricci scalar written in terms of the Levi-Civita connection. In teleparallel theories, the Ricci scalar is zero by definition, so since $R=0$, the Ricci scalar due to the Levi-Civita connection is related to $\mathbb{Q}$ as:
\begin{equation}
\mathcal{R}=-\mathbb{Q}-D_A(Q^A-\stackrel{_{-}}{Q}^A), 
\end{equation}

Thus, we can write the action for a non-metric gravity theory in terms of the non-metricity scalar as:
\begin{equation}
\label{açaoGTS}
S_{GTS}=-\dfrac{1}{16\pi G}\int d^4x\sqrt{-g}\mathbb{Q},
\end{equation}
and this is the action for Symmetric Teleparallel Gravity (STG). STG is a geometric description of gravity that is fully equivalent to general relativity. We can prove this equivalence in a straightforward manner in the so-called coincident gauge, where $\Gamma^{\sigma}_{\mu\nu}\equiv0$. Then, by imposing that the connection is symmetric, the torsion tensor is identically zero, and the Levi-Civita connection can be expressed in terms of the deformation tensor as \cite{BeltranJimenez:2018vdo}:
\begin{equation}
\label{gaugecoincident}\Gamma^{\sigma}_{\mu\nu}=-L^{\sigma}_{\mu\nu}.
\end{equation}

We can also work with a modified theory of symmetric teleparallel gravity. The idea is quite simple and follows a similar approach to those used in $f(R)$ \cite{DeFelice:2010aj} and $f(T)$ \cite{Cai:2015emx} theories. We will replace the non-metricity scalar in the action (\ref{açaoGTS}) with a function of the non-metricity scalar itself, which we will call $f(\mathbb{Q})$. Thus, the action for a modified theory of symmetric teleparallel gravity is given by \cite{Xu:2019sbp}:
\begin{equation}
\label{açãoGTS_Modificada}
S=\int d^Dx\sqrt{-g}\left( \frac{1}{2\kappa_D}f(\mathbb{Q})-2\Lambda+\mathcal{L}_{M} \right),
\end{equation}
where $\kappa_4=8\pi G_D$, with $G_D$ being the $D-$dimensional Newtonian gravitational constant, $\Lambda$
represents the bulk cosmological constant, and $\mathcal{L}_{M} $ is the matter-energy Lagrangian. Variation of the action (\ref{açãoGTS_Modificada}) with respect to the metric leads to the gravitational field equations \cite{Xu:2019sbp}
\begin{equation}
\label{equationsofcamp}
    \dfrac{2}{\sqrt{-g}}\nabla_K(\sqrt{-g}f_{\mathbb{Q}}P^K\ _{MN}) -\dfrac{\left(f-2\Lambda\right)}{2}g_{MN} + f_{\mathbb{Q}}(P_{MKL}Q_N\ ^{KL}-2Q_{KM}\ ^LP^K_{NL})=k_D \mathcal{T}_{MN}
\end{equation}
where $f_{\mathbb{Q}}\equiv\dfrac{df}{d\mathbb{Q}}.$ 
The energy-momentum tensor is given by,
\begin{equation}
  \mathcal{T}_{\mu\nu}\equiv -\dfrac{2}{\sqrt{-g}}\dfrac{\delta(\sqrt{-g}\mathcal{L}_{M})}{\delta g^{\mu\nu}} .
\end{equation}

\section{Evolution of the scale factor.}

In the following we will use the expressions of above in scenarios involving braneworld cosmology. We will consider a flat, homogeneous, and isotropic 5-dimensional spacetime, in which the metric takes the following form:
\begin{equation}
\label{ansatz}
ds^2=a^2(y)\left\{-dt^2+b^2(t)\hat{g}_{\mu\nu}dx^{\mu}dx^{\nu}\right\}+dy^2,
\end{equation}
where $\hat{g}_{\mu\nu}$ corresponds to the spatial components of the Friedmann-Lemaître-Robertson-Walker metric, and the coordinate ``y'' represents the extra dimension. The Greek indices $\mu$, $\nu$, ..., range from 1 to 3 and represent the spatial coordinates of the brane, while the capital Latin indices M, N, ..., represent the coordinates of the bulk and range from 0 to 4. The function $a(y)$  is the warp factor of the brane, while the function $b(t)$ is the scale factor.

Considering the coincident gauge ($\Gamma^{\rho}_{\mu\nu}=0$ \cite{Heisenberg:2023lru}), the non-zero components of the non-metricity tensor for the metric (\ref{ansatz}) are:
\begin{equation}
Q_{4\mu\nu}=2aa'b^2\hat{g}_{\mu\nu}, \ \ \ Q_{0\mu\nu}=2b\dot{b}a^2\hat{g}_{\mu\nu}, \ \ \ Q_{400}= -2aa',
\end{equation}
where a prime corresponds to the derivative with respect to $y$, and a dot corresponds to the derivative with respect to $t$. We also calculate the trace vectors:
\begin{equation}
Q_4=Q^4 =8\dfrac{a'}{a}, \ \ \ Q^0 =-6\dfrac{\dot{b}}{a^2b}, \ \ Q_0 =6\dfrac{\dot{b}}{b}.
\end{equation}
The non-metricity scalar in 5-dimensions is given by:
\begin{equation}
\label{nonmetricityscalar}
\mathbb{Q}=-4\left[3(c_1+3c_3)\dfrac{\dot{b}^2}{a^2b^2}-4(c_1+4c_3)\dfrac{a'^2}{a^2}\right],
\end{equation}
which depends on both the extra dimension $y$ and the cosmological time $t$. For constant $b(t)$, we obtain the same result as \cite{Silva:2022pfd}.

We wish to understand the gravitational effects along the extra dimension, so we propose a flat 3-brane whose source is a minimally coupled real scalar field with the following form:
\begin{equation}
\label{acao}
\mathcal{S}_{\mathcal{M}}=\int d^5x\sqrt{-g}\left(-\dfrac{1}{2} g^{AB}\nabla_A\phi\nabla_B\phi - V(\phi)\right),
\end{equation}
where $\phi \equiv \phi(y,t)$. The energy-momentum tensor for the scalar field $\phi(y,t)$ is given by:
\begin{equation}
\label{tensorenergiamomento}
T_{MN}=\nabla_M\phi\nabla_N\phi-g_{MN}\left(\dfrac{1}{2}(\nabla\phi)^2+V(\phi)\right).
\end{equation}

The equations of motion for the metric and scalar field from the actions (\ref{acao}) and (\ref{tensorenergiamomento}) are as follows:

(Component - 00):
\begin{equation}
\label{componente00}
\begin{split}
-4(c_1+4c_3)\left[f'_{\mathbb{Q}}\dfrac{a'}{a}+f_{\mathbb{Q}}\left(3\dfrac{a'^2}{a^2}+\dfrac{a''}{a}\right)\right]+ 6(c_5+2c_3)\left[ \dot{f_{\mathbb{Q}}}\dfrac{\dot{b}}{a^2b}+f_{\mathbb{Q}}\left( 2\dfrac{\dot{b}^2}{a^2b^2}+\dfrac{\ddot{b}}{a^2b}\right) \right]\\ + \dfrac{f}{2} -\Lambda^{(5)}+12f_{\mathbb{Q}} (c_1+3c_3)\dfrac{\dot{b}^2}{a^2b^2}= \dfrac{k_5}{2a^2}\dot{\phi}^2+\dfrac{k_5}{2}\phi'^2+k_5V(\phi)
\end{split}
\end{equation}
(Component - $\alpha,\beta={1,2,3}$):
\begin{equation}
\label{componenteab}
\begin{split}
-4(c_1+3c_3)\left[\dot{f_{\mathbb{Q}}}\dfrac{\dot{b}}{ba^2}+f_{\mathbb{Q}}\left(2\dfrac{\dot{b}^2}{b^2a^2}+\dfrac{\ddot{b}}{ba^2}\right)\right]+4(c_1+4c_3)\left[f'_{\mathbb{Q}}\dfrac{a'}{a}+f_{\mathbb{Q}}\left( 3\dfrac{a'^2}{a^2}+\dfrac{a''}{a}\right) \right]\\ -\dfrac{f}{2}+\Lambda^{(5)}\
=-\dfrac{k_5}{2}\phi'^2+\dfrac{k_5}{2a^2}\dot{\phi}^2-k_5V(\phi),
\end{split}
\end{equation}
(Component - 44):
\begin{equation}
\label{componente44}
\begin{split}
-12\dfrac{c_3}{a^2}\left[\dot{f_{\mathbb{Q}}}\dfrac{\dot{b}}{b}+f_{\mathbb{Q}}\left(2\dfrac{\dot{b}^2}{b^2}+\dfrac{\ddot{b}}{b}\right)\right] +8(c_5+2c_3)\left[f_{\mathbb{Q}}'\dfrac{a'}{a}+f_{\mathbb{Q}}\left(3\dfrac{a'^2}{a^2}+\dfrac{a''}{a}\right)\right]\\ -\dfrac{f}{2} +\Lambda^{(5)} +16f_{\mathbb{Q}}(c_1+4c_3)\dfrac{a'^2}{a^2}=\dfrac{k_5}{2}\phi'^2+\dfrac{k_5}{2a^2}\dot{\phi}^2-k_5V(\phi),
\end{split}
\end{equation}
where $\Lambda^{(5)}$ is a cosmological constant of the bulk.\\
(Scalar field):
\begin{equation}
\label{camposcalar}
\ddot{\phi}+3\dfrac{\dot{b}}{b}\dot{\phi}-4\phi'a'a+a^2V_{\phi}-a^2\phi''=0,
\end{equation}
where $V_{\phi}=\dfrac{dV}{d\phi}$.

\subsection{Static Solutions for quadratic gravity: $\kappa=0$}

We will now consider functions of the non-metricity scalar $\mathbb{Q}$ of the form:
\begin{equation}
\label{f(Q)}
    f(\mathbb{Q})= \mathbb{Q}+\kappa\mathbb{Q}^n,
\end{equation}
where $\kappa$ and $n$ are real numbers.
 Let us consider the case where $\kappa=0$, i.e., quadratic gravity. The equations (\ref{componente00}), (\ref{componenteab}), and (\ref{componente44}), in the static case $\phi=\phi(y)$, become:
\begin{align}
\label{00quadratica}
  2\sigma_0\left(\dfrac{\ddot{b}}{b}+2\dfrac{\dot{b}^2}{b^2}\right)+2\sigma_2\dfrac{\dot{b}^2}{b^2}&= \dfrac{a^2}{3}\left[4\sigma_1\left(\dfrac{a'^2}{a^2}+\dfrac{a''}{a}\right)+\dfrac{k_5}{2}\phi'^2+k_5V(\phi)+\Lambda^{(5)}\right],\\
\label{abquadratica}
    2\sigma_2\left(2\dfrac{\ddot{b}}{b}+\dfrac{\dot{b}^2}{b^2}\right)&=a^2\left[4\sigma_1\left(\dfrac{a'^2}{a^2}+\dfrac{a''}{a}\right)+\dfrac{k_5}{2}\phi'^2+k_5V(\phi)+\Lambda^{(5)}\right],\\
    \label{44quadratica}
    12c_3\left(\dfrac{\ddot{b}}{b}+2\dfrac{\dot{b}^2}{b^2}\right) -6\sigma_2\dfrac{\dot{b}^2}{b^2}&=a^2\left[8(3\sigma_0+\sigma_1)\dfrac{a'^2}{a^2}+8\sigma_0\dfrac{a''}{a}-\dfrac{k_5}{2}\phi'^2+k_5V(\phi)+\Lambda^{(5)}\right],
  \end{align}
and the equation for the scalar field becomes:
\begin{equation}
\label{campoescalarestacionario}
\phi''+4\dfrac{\phi'}{a}a'-V_{\phi}=0,
\end{equation}
where we define the following parameters: $\sigma_0\equiv c_5+2c_3$, $\sigma_1\equiv c_1+4c_3$, and $\sigma_2\equiv c_1+3c_3$.

For a specific choice of the parameters $c_i$, symmetric teleparallelism becomes equivalent to general relativity. This choice is given by \cite{Heisenberg:2023lru}:
\begin{equation}
\label{Escolha_coeficientes}
\boxed{c_1=-c_3=-\dfrac{1}{4}, \ \ \ c_2=-c_5=\dfrac{1}{2}, \ \ c_4=0,}
\end{equation}
When adopting this choice, we work within the general relativity regime. Therefore, in this regime we obtain  $\sigma^{RG}_0=0$, $\sigma^{RG}_1=3/4$, $\sigma^{RG}_2=1/2$ and equations (\ref{00quadratica}), (\ref{abquadratica}), and (\ref{44quadratica}) become the Einstein's equations for a thick brane in the usual general relativity, for $k=0$ (spatially flat universe)  \cite{Ahmed:2013lea, Hendi:2020qkk}. It is also interesting to note that equations (\ref{componenteab}) and (\ref{componente44}), for a warp factor $a(y)=e^{A(y)}$, and for constant $b(t)$, become the equations of motion of the action of a thick flat 3-brane in $f(\mathbb{Q})$ gravity \cite{Silva:2022pfd}.

Combining equations (\ref{00quadratica}) - (\ref{44quadratica}), we obtain this equation:
\begin{equation}
 \label{FriedmannModif2}
(6c_3+3\sigma_0+2\sigma_2)\dfrac{\ddot{b}}{b}+(6\sigma_0+2\sigma_2+12c_3)\dfrac{\dot{b}^2}{b^2}=\ \Lambda(y), 
 \end{equation}
where the term $\Lambda(y)$ is obtained by transposing all terms that depend on the extra dimension in equation (\ref{FriedmannModif2}) to the right side of the equation, and is therefore given by
 \begin{equation}
\label{Constante_Cosmologica_effetiva.}
     \Lambda(y) = 4[(2\sigma_1+3\sigma_0)a'^2+(\sigma_0+\sigma_1)a''a]+\dfrac{a^2}{4}\phi'^2+\dfrac{3a^2}{2}(V(\phi)+\Lambda^{(5)}), 
 \end{equation} 
 where was used $k_5=1$.
  
Defining the Hubble factor as $H\equiv\frac{\dot{b}(t)}{b(t)}$, we can rewrite equation (\ref{FriedmannModif2}) as
\begin{equation}
\label{FriedmannModif}
A\dot{H}+(A+B)H^2=\Lambda(y),
\end{equation}
where $A=6c_3+3\sigma_0+2\sigma_2$ and $B=6\sigma_0+2\sigma_2+12c_3$. Let us treat $\Lambda(y)$ as a constant for now, when we can find some solutions that depend on the sign of $\Lambda(y)$. The solutions are in the table below.

\begin{table}[h!]
\centering
\footnotesize
\caption{Solutions of the Modified Friedmann Equation and Universe Types}
\label{tab:friedmann_solutions}
\begin{tabular}{lcc}
\hline
\textbf{Parameter Conditions} & \textbf{Solution for the Scale Factor \( b(t) \)} & \textbf{Universe Type} \\
\hline
\( (1)\ \Lambda = 0,\ \ A+B>0 \) & 
\( b(t) = b_0 \left| \dfrac{A+B}{A} t - C \right|^{\frac{A}{A+B}} \) & 
Power-law expansion/contraction \\ 
\((2)\ \Lambda > 0,\ \ A+B>0 \) & 
\( b(t) = b_0 \left[ \cosh\left( k \dfrac{A+B}{A} t + C \right) \right]^{\frac{A}{A+B}} \) & 
Accelerated expansion (de Sitter type) \\ 
\( (3)\ \Lambda < 0,\ \ A+B>0 \) & 
\( b(t) = b_0 \left| \cos\left( k \dfrac{A+B}{A} t - C \right) \right|^{\frac{A}{A+B}} \) & 
Oscillatory universe \\ 
\((4)\Lambda >0,\ A = 0 \) & 
\( b(t) = b_0 \exp\left( \pm \sqrt{\dfrac{\Lambda}{B}} t \right) \) & 
Exponential expansion/contraction \\ 
\((5)\Lambda >0,\ A + B = 0 \) & 
\( b(t) = b_0 e^{\pm H t} \) & 
Exponential expansion/contraction \\ 
\hline
\end{tabular}
\end{table}
 where: \( A = 6c_3 + 3\sigma_0 + 2\sigma_2 \), \( B = 6\sigma_0 + 2\sigma_2 + 12c_3 \), \( k = \sqrt{\frac{|\Lambda|}{A+B}} \), and \( b_0 \) is the initial scale factor and \( C \) is an integration constant. The function $b(t)$ is defined for a specific value of $y$, and therefore will depend on time and the extra dimension. Let us consider that for $y=0$ it reduces to the scale factor of the brane. We therefore see that $\Lambda(y)$ closely resembles a cosmological constant, although it is not really a constant, since it is still a function of the extra dimension.


Expressing the warp factor as $a(y)\equiv e^{A(y)}$, allows us to write equation (\ref{Constante_Cosmologica_effetiva.}) as:
\begin{equation}
\label{Eq:Cosntantecosmologicaefetiva}
    \Lambda(y) = 4[(3\sigma_1+2\sigma_0)A'^2+(\sigma_0+\sigma_1)A'']e^{2A}+\dfrac{3}{2}(V(\phi)+\Lambda^{(5)})e^{2A}+\dfrac{1}{4}\phi'^2e^{2A}, 
\end{equation}
 As demonstrated, this quantity depends exclusively on the extra dimension. The solutions listed in Table \ref{tab:friedmann_solutions} indicate that $\Lambda(y)$ behaves as an effective cosmological constant in the bulk. Since we locate the brane at the origin of the coordinate system, by setting $y=0$, we obtain an effective cosmological constant defined on the brane, which we express as $\Lambda^{(4)} = \Lambda(0)$.

Our objective, therefore, is to study equation (\ref{Eq:Cosntantecosmologicaefetiva}) to determine how $A(y)$, $\phi$, $V(\phi)$ and $\Lambda^{(5)}$ collectively determine the sign of $\Lambda^{(4)}$ and, consequently, which type of solution is most suitable for the brane scale factor. To this end, we will propose two configurations: one involving a thin brane and another a thick brane, in which the functions associated with the extra dimension will assume a specific form.

\subsubsection{Thin brane}

\begin{figure}
\centering
\includegraphics[width=90mm]{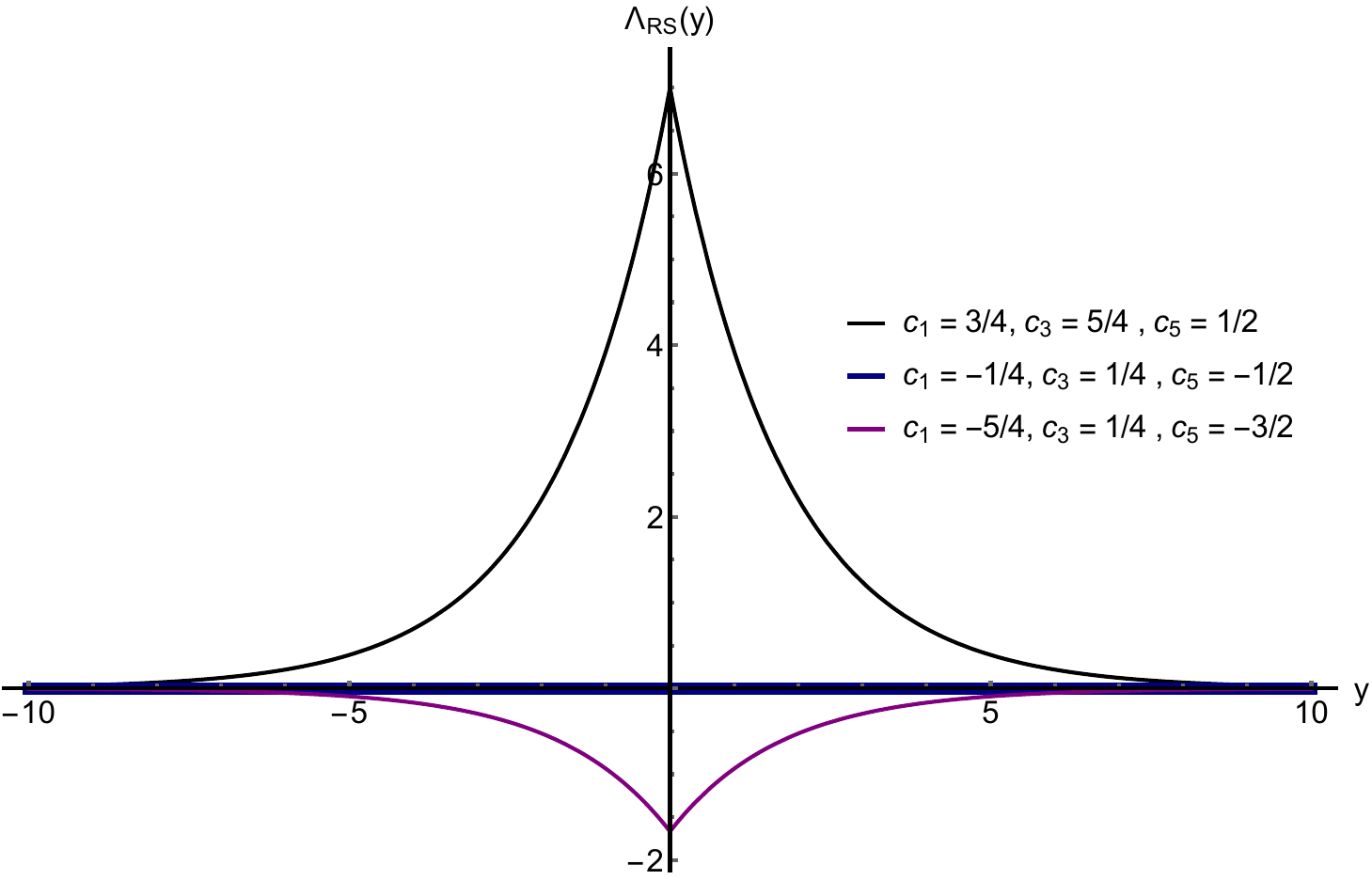}
\caption{$\Lambda(y)$ for RS solution ( \ref{lambdaRS}) for different values of the parameters $c_i$. }
\label{fig:modeloRS}
\end{figure}

\begin{figure}[!htb]
\centering
\includegraphics[width=80mm]{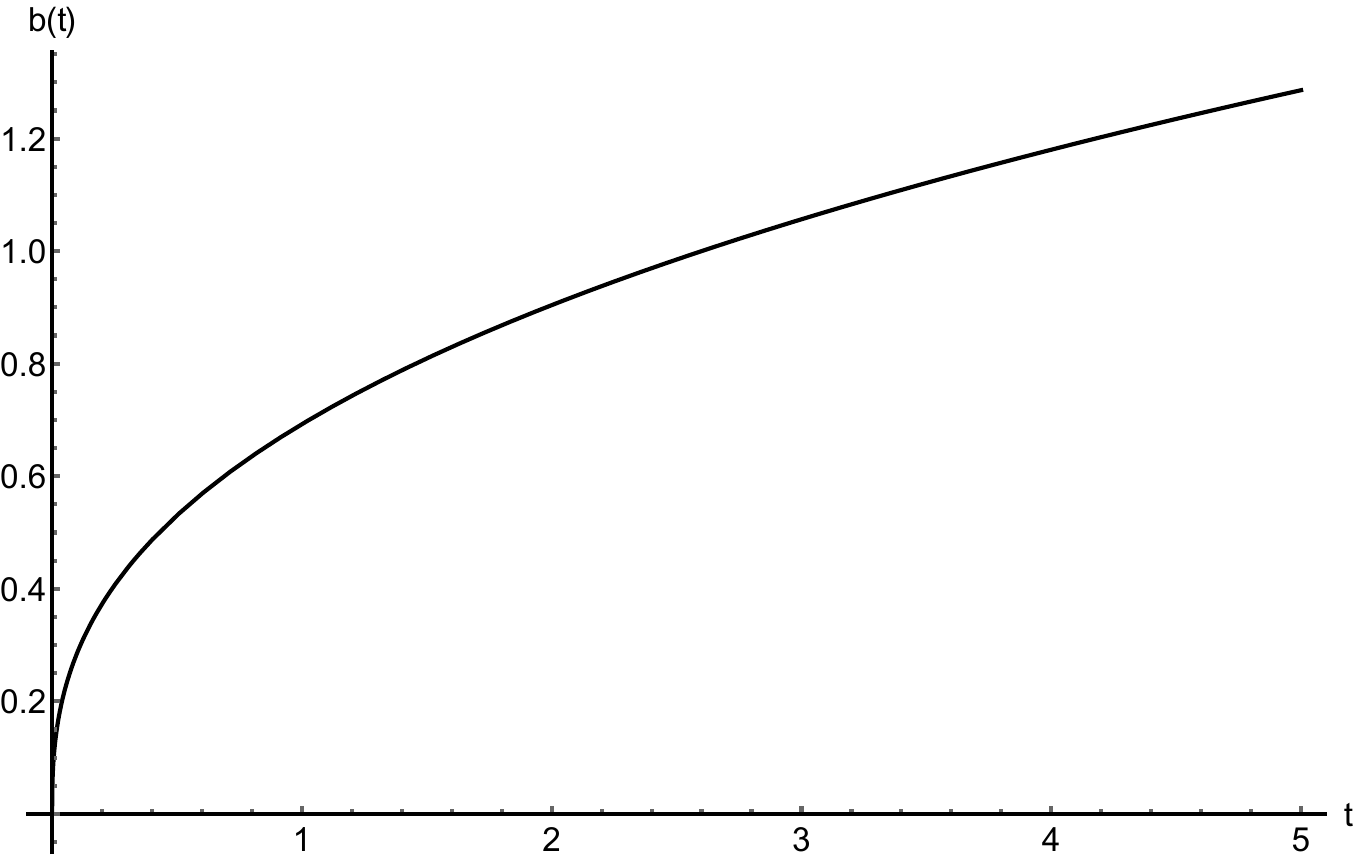}
\caption{Scale factor for RS solution (\ref{b(t)RS1}), for $b_0=1$, $\frac{A}{A+B}=5/13$ and $C=0.$ }
\label{fig:b(t)RS}
\end{figure}

We will consider the  case of the thin brane.
Proposing an RS (Randall-Sundrum) solution for $A(y)$,
\begin{equation}
    A(y)=-\sqrt{\frac{-k_5^2}{6}\Lambda^{(5)}}|y|\equiv -k|y|
\end{equation}
where $\Lambda^{(5)}$ is the cosmological constant of bulk. Let us also take that
$\phi=V(\phi)=0$.
We have:
\begin{equation}
\Lambda_{RS}(y)=-\frac{4}{6}(3\sigma_1+2\sigma_0)\Lambda^{(5)}e^{-2k|y|}   +\dfrac{3}{2}\Lambda^{(5)}e^{-2k|y|},
\end{equation}
 Thus, by writing $\Lambda(y=0)=\Lambda^{(4)}$, on the brane we have that
  \begin{equation}
  \label{lambdaRS}
\Lambda^{(4)}_{RS}=-\frac{2}{3}(3\sigma_1+2\sigma_0)\Lambda^{(5)}+\dfrac{3}{2}\Lambda^{(5)}.
\end{equation}
The requirement that $\Lambda^{(4)}>0$, and that the bulk remains  Anti-deSitter($AdS_5$), leads us to:
\begin{equation}
\label{fine_turnig_Lambda}
  \boxed{  3\sigma_1+2\sigma_0>9/4}.
\end{equation} 
Furthermore, according to (\ref{lambdaRS}), if $3\sigma_1+2\sigma_0=9/4$, then $\Lambda^{(4)}=0$, regardless of the value of $\Lambda^{(5)}$. If we choose the values of the parameters $c_i$ from the general relativity limit (choice given by (\ref{Escolha_coeficientes})), we obtain a null value for $\Lambda^{(4)}$. In this case, the scale factor solution falls into the first case of Table \ref{tab:friedmann_solutions}:
\begin{equation}
\label{b(t)RS1}
b(t) = b_0 \left| \dfrac{A+B}{A} t - C \right|^{\frac{A}{A+B}}.
\end{equation}
We then choose $C=0$ for the integration constant, and in the GR regime we have $\frac{A}{A+B}=\frac{5}{13}$; the obtained solution is shown in Figure \ref{fig:b(t)RS}. Ou seja, embora a constante cosmológica seja nula ainda assim obtemos uma soulução de universo em expansão, embora seja uma espansão bem lenta.

\begin{figure}[!htb]
\centering
\includegraphics[width=80mm]{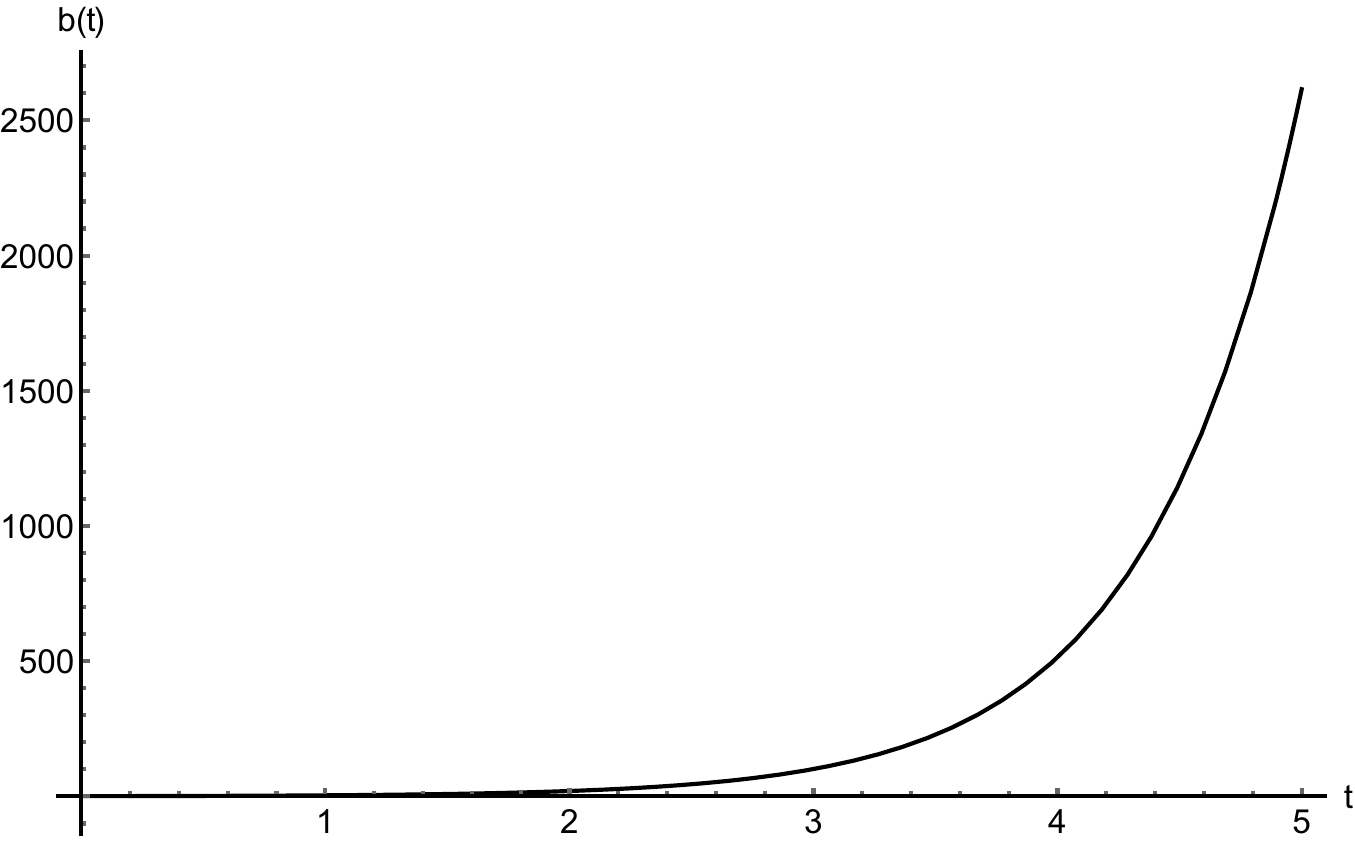}
\caption{Scale factor for RS solution (\ref{b(t)RS2}), for $b_0=1$, $c_1=3/4$, $c_3=5/4$, $c_5=1/2$ and $C=0.$ }
\label{fig:b(t)RSNRG}
\end{figure}

If we choose values for the parameters $c_i$ such that $3\sigma_1+2\sigma_0>9/4$, we obtain de Sitter-type solutions for the brane, while the bulk remains Anti-de Sitter. The solution will now be the second equation in Table \ref{tab:friedmann_solutions}, which we can write as:
\begin{equation}
\label{b(t)RS2}
b(t) = b_0 \left[ \cosh\left( \dfrac{\sqrt{\Lambda^{(4)}_{RS}(A+B)}}{A} t + C \right) \right]^{\frac{A}{A+B}}.
\end{equation}
We plot the scale factor for this case in Figure \ref{fig:b(t)RSNRG}.

\begin{figure}[!htb]
\centering
\includegraphics[width=80mm]{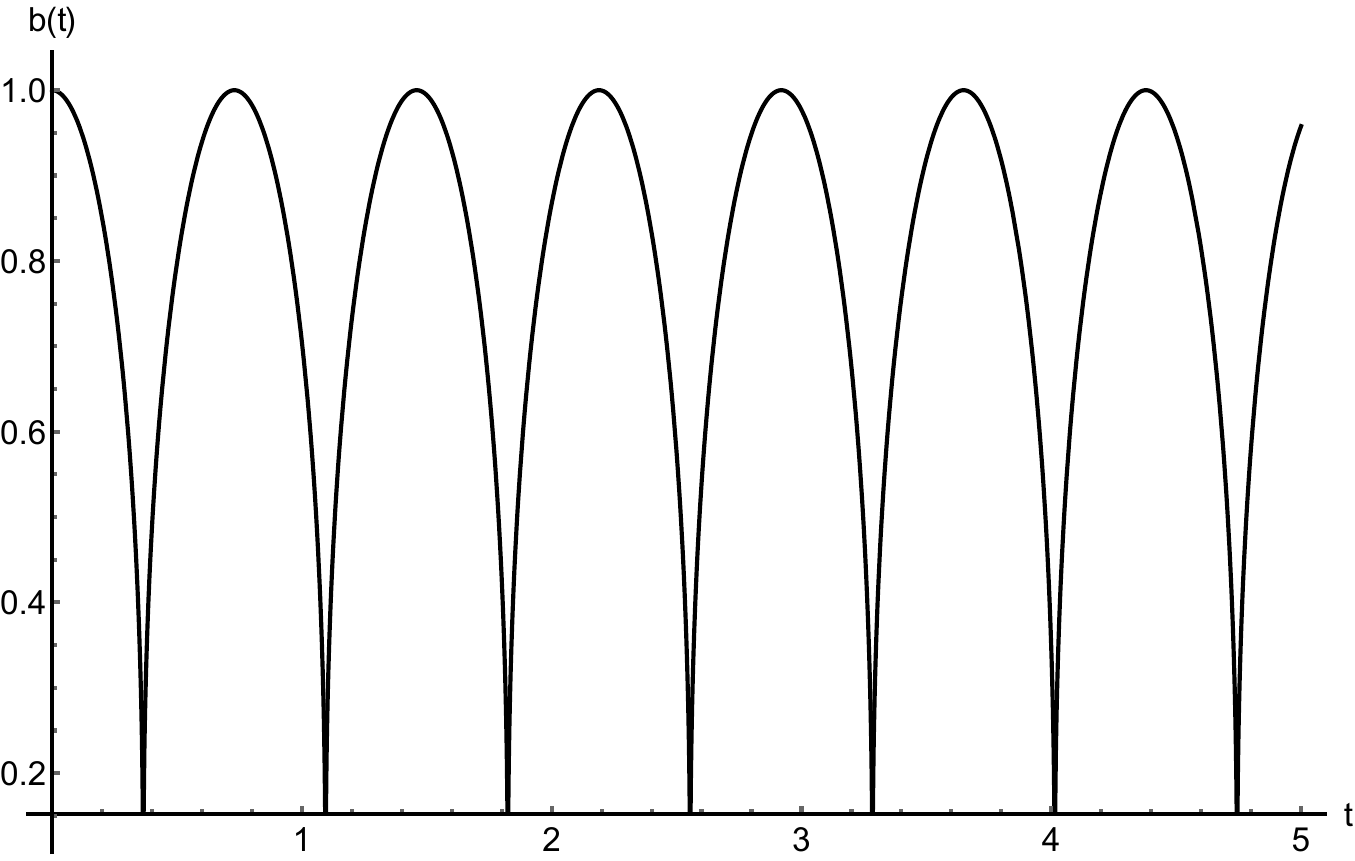}
\caption{Scale factor for RS solution (\ref{b(t)RS3}), for $b_0=1$, $c_1=-5/4$, $c_3=1/4$, $c_5=-3/2$ and $C=0.$ }
\label{fig:b(t)rs3}
\end{figure}

To complete the possible cases, we will choose values for the parameters $c_i$ for which the cosmological constant is negative. A possible choice would be $c_1=-5/4$, $c_3=1/4$ and $c_1=-3/2$, and we would still have $A+B>0$. The solution for the scale factor is given by
\begin{equation}
\label{b(t)RS3}
b(t) = b_0 \left[ \cos\left( \dfrac{\sqrt{\Lambda^{(4)}_{RS}(A+B)}}{A} t -C \right) \right]^{\frac{A}{A+B}},
\end{equation}
which is the case of an oscillatory universe; we represent this solution in figure \ref{fig:b(t)rs3}.

Some observations regarding the obtained results are necessary. Firstly, our solutions demonstrate that the introduction of an extra dimension generates expanding cosmological scenarios, including an expansion regime characteristic of a universe with a nonzero cosmological constant, a behavior consistent with current cosmological observations. Furthermore, as illustrated in Figure \ref{fig:modeloRS}, the curvature of the bulk confines the effective cosmological constant near the origin of the system, where it assumes its maximum absolute value, decaying asymptotically to zero as we move away from the origin. This implies that if the brane were sufficiently far from the origin, the effective cosmological constant would be null. A conceptually interesting possibility then arises: in the context of the Randall-Sundrum type I model \cite{Randall:1999ee}, which features two branes, a configuration would be possible in which one brane, located at the origin, would exhibit a nonzero cosmological constant, while the other, situated at a sufficient distance, would have a null effective cosmological constant.

\subsubsection{Thick brane}

Considering the equation (\ref{Effective_Cosmological_Constant}) for the case of a thick brane. The left side of the equations (\ref{00quadratica}), (\ref{abquadratica}), and (\ref{44quadratica}) provide a system of equations for the warp factor $A(y)$ , the scalar field $\phi(y)$, and the potential $V(\phi)$ in the extra dimension , if we know how to solve them we would know how to relate these quantities with the cosmological constant in the bulk in order to know the sign of the cosmological constant in the brane. Fortunately, we can adopt a first-order formalism by writing the warp factor and the scalar field as \cite{Afonso:2006gi,Gremm:1999pj}:
\begin{equation}
\label{BPS_A}
    A'=-\frac{1}{3}W,
\end{equation}
\begin{equation}
\label{BPS_phi}
\phi'=\sqrt{2\alpha}\dfrac{\partial W}{\partial \phi},
\end{equation}
where $W(\phi)$ is the superpotential.
Now the potential now has the form:
\begin{equation}
\label{BPS_V}
    V(\phi)=\alpha\left(\dfrac{\partial W}{\partial \phi}\right)^2-\dfrac{1}{3}W^2,
\end{equation}
where $\alpha=(2\sigma_0+\sigma_1)/6$.




We will use the Sine-Gordon model as used in \cite{Silva:2022pfd}, where the superpotential is given by:
\begin{equation}
\label{sine_Gordon_superpotential}
    W(\phi)=3bd \sin\left[ 
\sqrt{\dfrac{2}{3b}} \phi\right],
\end{equation}
where $d$ and $b>0$ are parameters.
Solve the equation (\ref{BPS_V}) we have a modified sine-Gordon potential: 
\begin{equation}
    V(\phi)=\dfrac{3bd^2}{2}\left((2\alpha-b)+(2\alpha+b)\right)\cos{\left(2\sqrt{\dfrac{2}{3b}}\phi\right)}.
\end{equation}
The scalar field solution of the BPS equation (\ref{BPS_phi}) leads to
\begin{equation}
\label{BPS_phi_soluction}
    \phi(y)= \sqrt{6b}\arctan{(\tanh{(\sqrt{2\alpha}dy}))}.
\end{equation}
And finally, solving equation (\ref{BPS_A}) we obtain the warp factor
\begin{equation}
    A(y)=ln(sech{(\lambda y)^{\frac{b}{\sqrt{8\alpha}}}}).
\end{equation}

The equation (\ref{Effective_Cosmological_Constant}) becomes, 
\begin{equation}
\label{Lambda(phi)}
    \Lambda(y)=\left(\dfrac{4(3\sigma_1+2\sigma_0)}{9}-\dfrac{1}{2}\right)W^2e^{2A(y)}+\left(2\alpha-\dfrac{2(\sigma_0+\sigma_1)\sqrt{\alpha}}{3} \right)\left(\dfrac{\partial W}{\partial \phi}\right)^2e^{2A(y)}
    \\+\dfrac{3}{2}\Lambda^{(5)}e^{2A(y)}.\nonumber
\end{equation}
The necessary condition for $\Lambda(y)$ to be real is that we have $\sigma_0$ and $\sigma_1$ such that
\begin{equation}
   \boxed{ 2\sigma_0+\sigma_1\ge0.}
\end{equation}

Let's now replace (\ref{sine_Gordon_superpotential}) in (\ref{Lambda(phi)}), and let $\Lambda^{(5)}=0$, then
\begin{align}
\label{lambdayGordon}
    \Lambda(y)=&9b^2d^2\left(\dfrac{4(3\sigma_1+2\sigma_0)}{9}-\dfrac{1}{2}\right)\sin\left[ 
\sqrt{\dfrac{2}{3b}} \phi\right]^2e^{2A(y)}\\&+6bd^2\left(2\alpha-\dfrac{2(\sigma_0+\sigma_1)}{3}\sqrt{\alpha} \right)\cos\left[ 
\sqrt{\dfrac{2}{3b}} \phi\right]^2e^{2A(y)}+\dfrac{3}{2}\Lambda^{(5)}e^{2A(y)},\nonumber
\end{align}
and  when  $\phi=0$, we have that $\Lambda(0)=\Lambda^{(4)}$, where
\begin{align}
    \Lambda^{(4)}=6bd^2\left(2\alpha -\dfrac{2(\sigma_0+\sigma_1)}{3}\sqrt{\alpha}\right)+\frac{3}{2}\Lambda^{(5)}.
\end{align}
The condition for $\Lambda^{(4)}$ to be positive when $\Lambda^{(5)}<0$ is that 
\begin{align}
    \boxed{(2\sigma_0 + \sigma_1) - \frac{2}{3}(\sigma_0 + \sigma_1) \sqrt{\frac{2\sigma_0 + \sigma_1}{6}} > \frac{-\Lambda^{(5)}}{4 b d^2}}
\end{align}
where $ 2\sigma_0 + \sigma_1 \ge 0$.

\begin{figure}[]
\centering
\includegraphics[width=80mm]{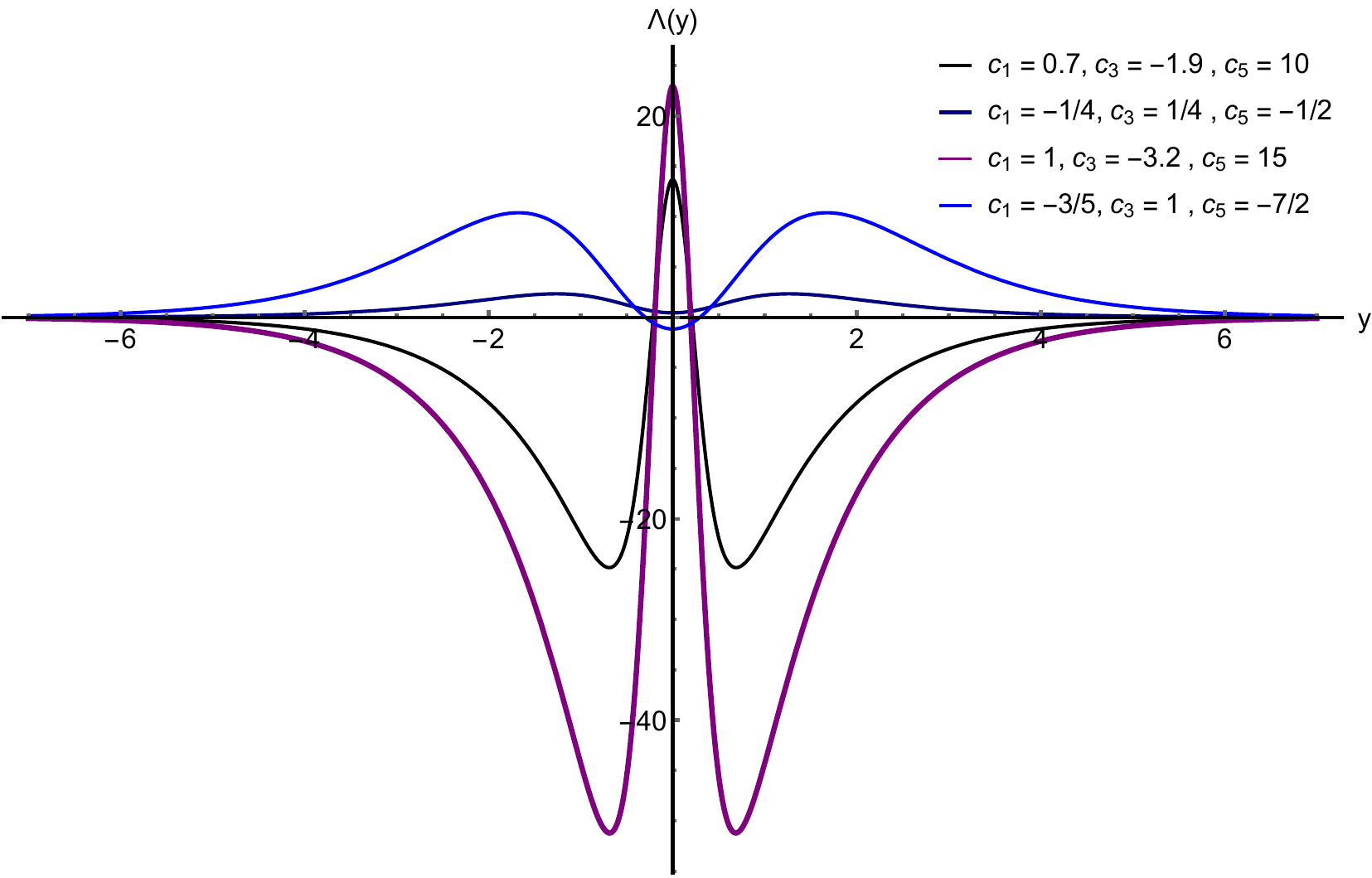}
\caption{$\Lambda(y)$ of (\ref{lambdayGordon}) for different values of the coefficients $c_i$, with $\Lambda^{(5)}=0$.}
\label{fig:campo_escalar}
\end{figure}

As we can see, we managed to obtain a positive cosmological constant on the brane even in the GR limit, $\sigma_0^{GR}=0$ and $\sigma_1^{GR}=3/4$, which was not possible in the RS solution. The behavior of $\Lambda(y)$ is quite interesting; for an appropriate choice of coefficients, we can have positive, negative, or null values for the effective cosmological constant on the brane, therefore the different solutions in table \ref{tab:friedmann_solutions} can be obtained.

Once again, the effective cosmological constant goes to zero asymptotically, but for certain choices of parameters, we can obtain values of the cosmological constant in the bulk that are different from those on the brane. For example, for the choice $c_1=0.7$, $c_1=-1.9$ and $c_1=10$, we have a positive value of the cosmological constant on the brane, but in the bulk, it will have negative values near the brane before going to zero as y increases.
This means that once again the curved space of the bulk tends to confine the cosmological constant near the origin, canceling it asymptotically. And again we see that the position of the brane in the bulk interferes with its cosmology, because the position it occupies in the bulk determines the value of the cosmological constant.  
\begin{figure}[]
\centering
\includegraphics[width=80mm]{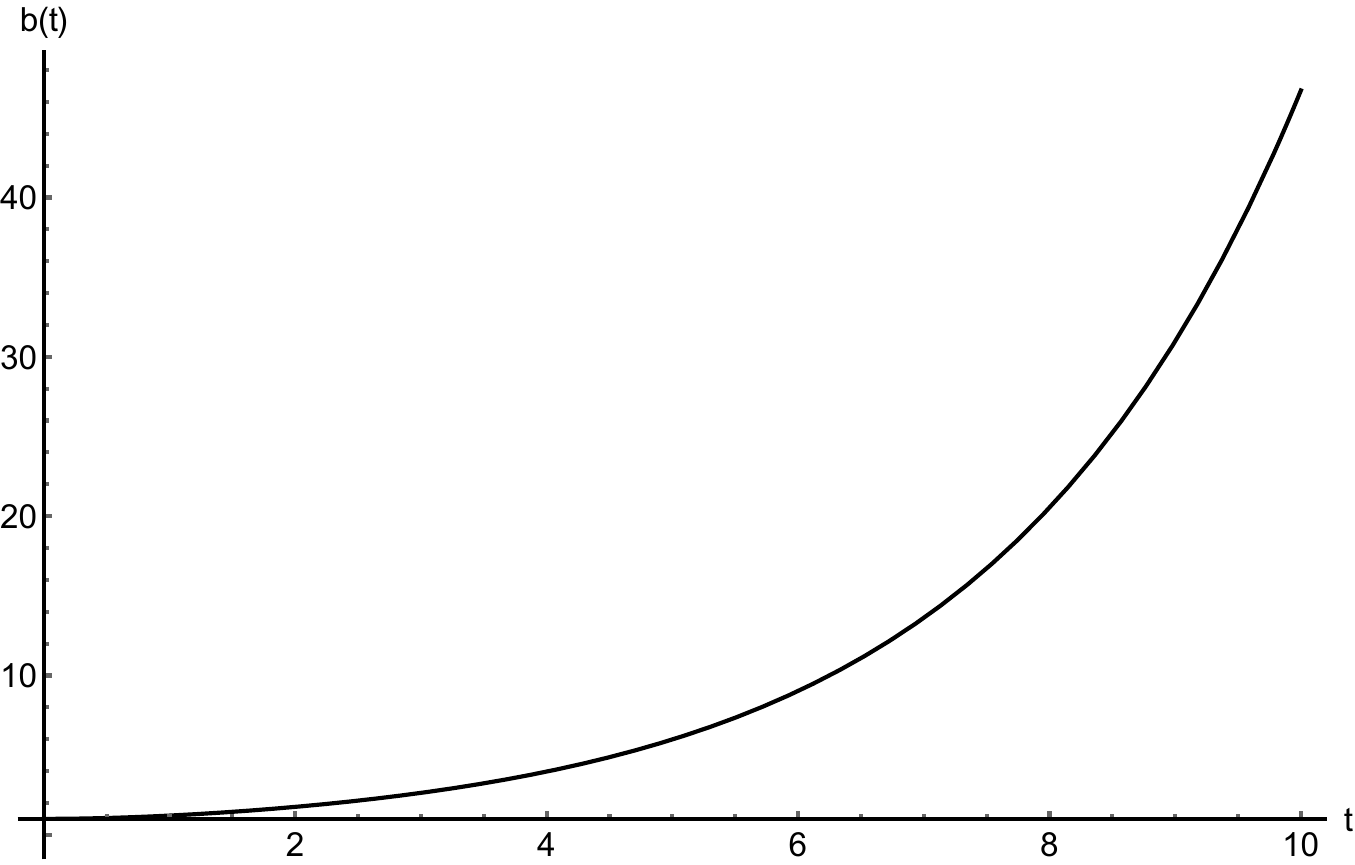}
\caption{Scale factor for a thick brane in the GR regime: $c_1=-1/4$, $c_3=1/4$ and $c_5=-1/2$.
}
\label{fig:campo_escalar2}
\end{figure}

\subsection{Static Solutions for $\kappa\ne0$}
We can also consider generalizations of the function $f(\mathbb{Q})$, defined in (\ref{f(Q)}), particularly for $\kappa \ne 0$ and $n > 1$. However, such generalizations introduce very complicated combinations involving cross terms between the scale and warp factors. Given these difficulties, we restrict our analysis to the case $n = 1$, which still corresponds to quadratic gravity.

Thus, for $\kappa \ne 0$ and $n = 1$ in (\ref{f(Q)}), we obtain the following modified Friedmann equation: 
\begin{equation}
        (6c_3+3\sigma_0+2\sigma_2)\dfrac{\ddot{b}}{b}+(6\sigma_0+2\sigma_2+12c_3)\dfrac{\dot{b}^2}{b^2}=\frac{ \Lambda_{\kappa}(y)}{\kappa+1},
\end{equation}
where
\begin{equation}
\label{Effective_Cosmological_Constant}
    \Lambda_{\kappa}(y) = 4(\kappa+1)[(3\sigma_1+2\sigma_0)A'^2+(\sigma_0+\sigma_1)A'']e^{2A}+\dfrac{3}{2}(V(\phi)+\Lambda^{(5)})e^{2A}+\dfrac{1}{4}\phi'^2e^{2A}.
\end{equation}
\begin{table}[h]
\centering
\footnotesize
\caption{Solutions of the modified Friedmann equation in symmetric teleparallel gravity}
\label{tab:tabela2}
\begin{tabular}{lcc}
\hline
\textbf{Parameter conditions} & \textbf{Scale factor solution } b(t) & \textbf{Universe type} \\
\hline
$(1)\ \Lambda = 0, A+B > 0$ & $b(t) = b_0 \left| \dfrac{A+B}{A} t - C \right|^{\frac{A}{A+B}}$ & Power-law expansion/contraction \\
$(2)\ \Lambda > 0$, $A+B > 0$ & $b(t) = b_0 \left[ \cosh\left( k \dfrac{A+B}{A} t + C \right) \right]^{\frac{A}{A+B}}$ & Accelerated expansion \\
$(3)\ \Lambda < 0$, $A+B > 0$ & $b(t) = b_0 \left| \cos\left( k \dfrac{A+B}{A} t - C \right) \right|^{\frac{A}{A+B}}$ & Oscillatory universe \\
$(4)\ A = 0$, $\Lambda > 0$ & $b(t) = b_0 \exp\left( \pm \sqrt{\dfrac{\Lambda}{B(\kappa+1)}} t \right)$ & Exponential expansion \\
\((5)\ \Lambda >0,\ A + B = 0 \) & 
\( b(t) = b_0 e^{\pm H t} \) & 
Exponential expansion/contraction \\
\hline
\end{tabular}
\end{table}
\noindent 
The solutions for the scale factor are presented in Table \ref{tab:tabela2}, where $A = 6c_3 + 3\sigma_0 + 2\sigma_2$, $B = 6\sigma_0 + 2\sigma_2 + 12c_3$, $k = \sqrt{\frac{|\Lambda|}{(A+B)(\kappa+1)}}$,  $b_0$ is the initial scale factor and $C$ is an integration constant. 

The behavior of $\Lambda_{\kappa}(y)$ in diferent cases is shown in the figures below.
\begin{figure}[!ht]
\centering
\includegraphics[width=120mm]{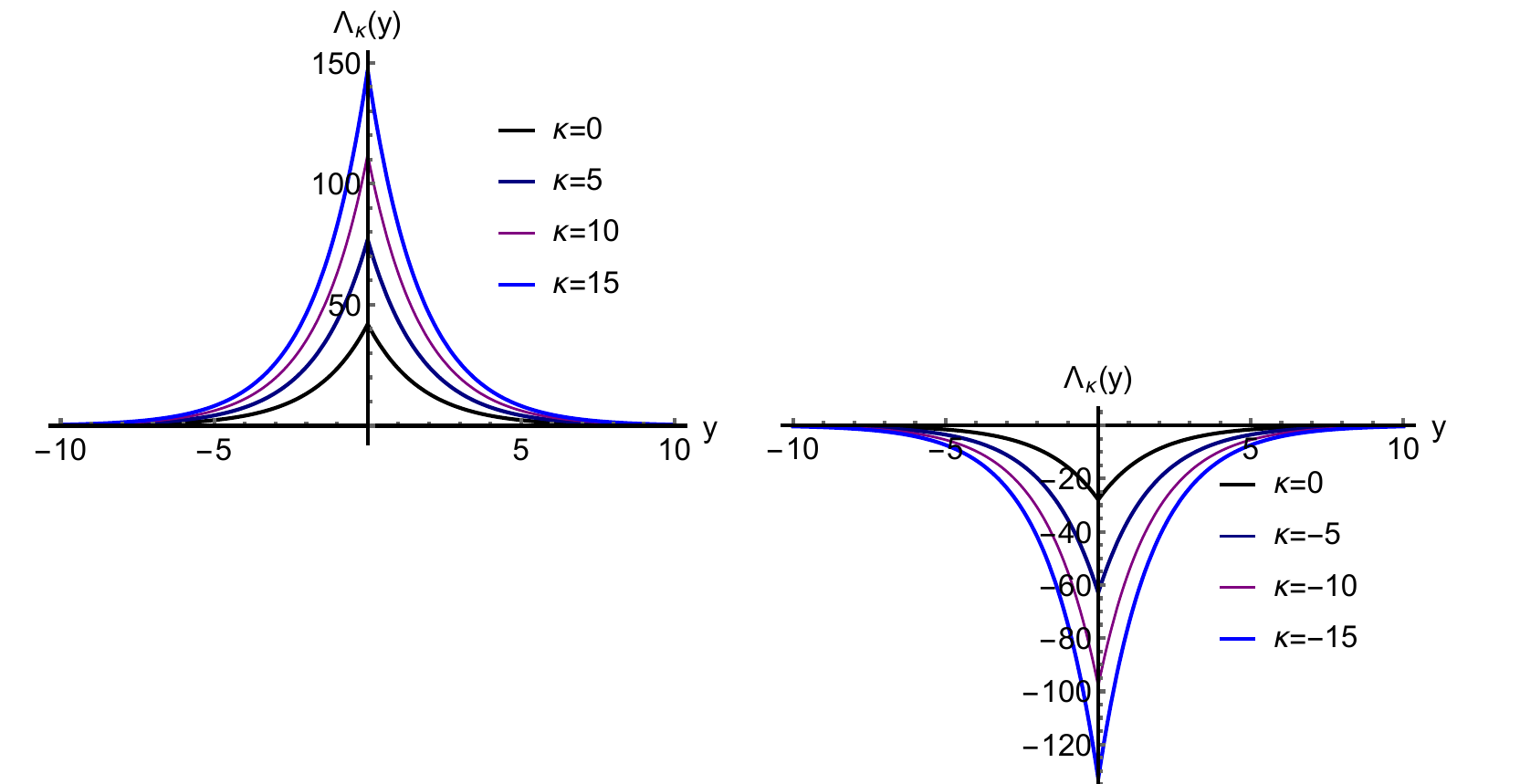}
\caption{$\Lambda_{\kappa}(y)$, for RS regime, for $c_1=3/4,\ c_3=5/4,\ c_5=1/2,$ with different values of $\kappa$.}
\label{fig:kappa_variando1}
\end{figure}
We verify that the presence of the $\kappa$ parameter introduces modifications to the effective cosmological constant. Depending on the sign and magnitude of this parameter, it is possible to alter the sign of the cosmological constant on the brane and model its behavior in the bulk. However, the cosmological constant maintains its asymptotic decay. Furthermore, it is observed that the confinement of the cosmological constant near the brane can be enhanced.
\begin{figure}[!ht]
\centering
\includegraphics[width=120mm]{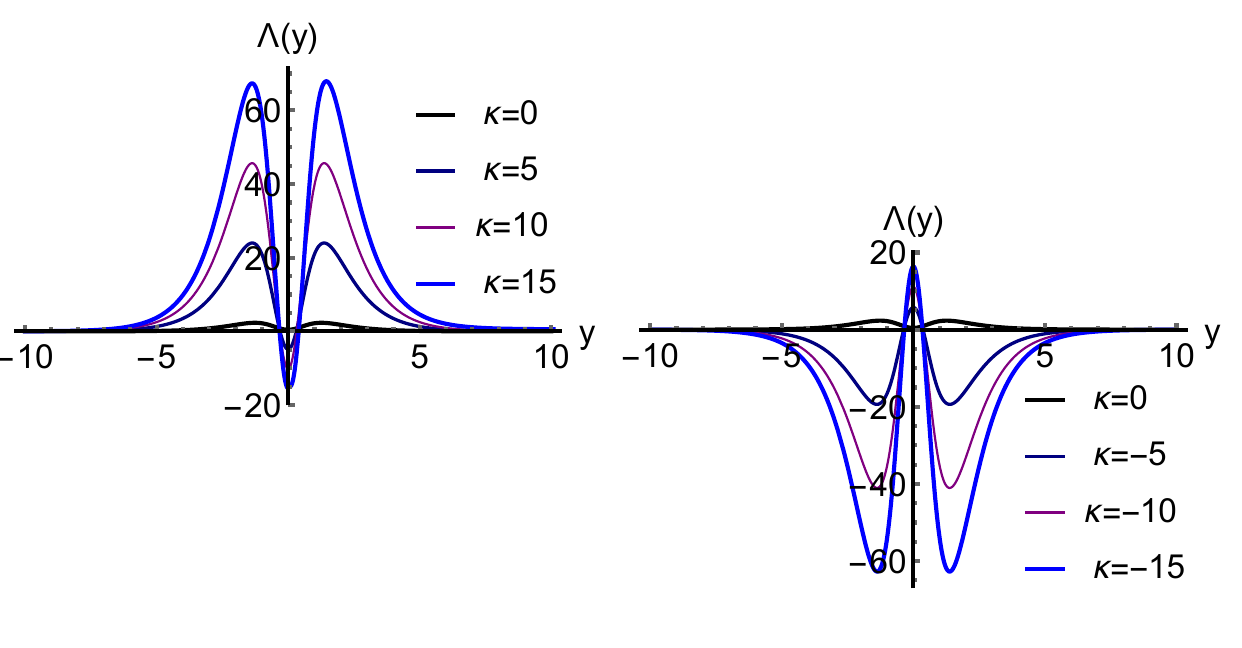}
\caption{$\Lambda_{\kappa}(y)$, for thick brane regime, for $c_1=-1/4,\ c_3=1/4,\ c_5=-1/2,$ with different values of $\kappa$.}
\label{fig:kappa_variando2}
\end{figure}
\begin{figure}[!ht]
\centering
\includegraphics[width=120mm]{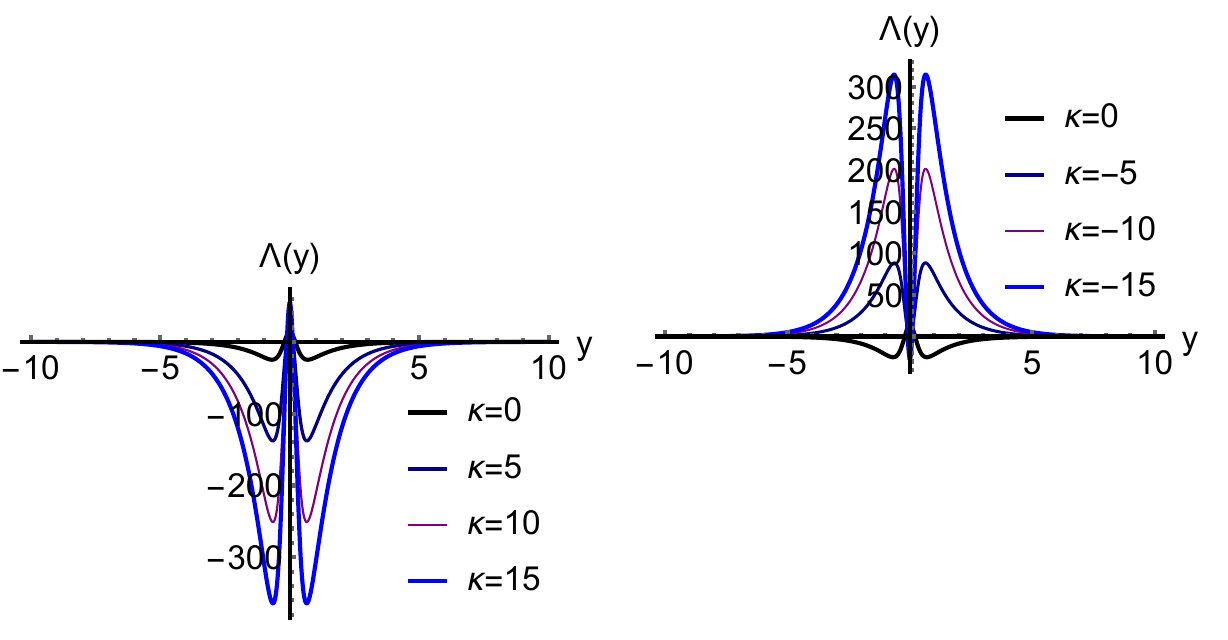}
\caption{$\Lambda_{\kappa}(y)$, in thick brane regime, for $c_1=0.7,\ c_3=-1.9,\ c_5=10,$ with different values of $\kappa$.}
\label{fig:kappa_variando3}
\end{figure}

\newpage


\section{Final remarks and perspectives}

In this work we studied a brane world cosmology model where gravitational dynamics are governed by a modified symmetric teleparallel gravitational theory. We considered an FLRW geometry for the brane immersed in a bulk filled with a scalar field that depend on the extra dimension. We have thus developed a more comprehensive theoretical framework compared to previous studies, which were limited to demonstrating that the extra dimension introduces an effective cosmological constant on the brane through a separation constant when solving Einstein's equations. We demonstrate that the effective cosmological constant can be treated as a function of the extra dimension, and that the curved bulk geometry not only generates this constant but also confines it near the origin of the coordinate system, inducing an asymptotic decay to zero. Thus, the brane's cosmology is determined by its position in the bulk.

The application of symmetric teleparallelism in the context of $f(\mathbb{Q})$ gravity has also yielded significant advances. Initially, we allowed the parameters $c_i$, which arise from the definition of the non-metricity scalar $\mathbb{Q}$, to assume arbitrary values. Variation of these parameters induced substantial modifications in the brane cosmology, enabling the derivation of multiple distinct solutions.

In the general relativity limit, corresponding to the specific parameter choice given in (\ref{Escolha_coeficientes}) where symmetric teleparallelism becomes equivalent to general relativity, we obtained particular solutions. In the thin brane regime, the effective cosmological constant on the brane is null; however, the solution of the modified Friedmann equation still guarantees an expanding universe, albeit at a reduced rate.  
In the context of a thick brane regime, we employ a first-order formalism in which the cosmological constant on the brane is determined by a superpotential, itself being a function of the field $\phi$. By utilizing the superpotential derived from the Sine-Gordon model, the effective cosmological constant of the brane is dictated by both the parameters of the Sine-Gordon model and the coefficients $c_i$ of symmetric teleparallelism.
In this context, within the same GR limit, the effective cosmological constant assumes positive values, though of moderate magnitude: it exhibits a slight increase when moving away from the origin, followed by an asymptotic decay to zero. This provides an explanation for why our cosmological constant is small, though non-zero.

Several promising perspectives thus emerge. In future work, we could explore the variation of the $c_i$ parameters in models where these parameters evolve with the universe, similar to models with time-dependent fundamental constants. There also exists the possibility of considering solutions where the scalar field depends on both time and the extra dimension, which might allow investigation of brane motion within the bulk and how this determines its cosmology.



\begin{thebibliography}{99}  

\bibitem{SupernovaSearchTeam:1998fmf}
A.~G.~Riess \textit{et al.} [Supernova Search Team],
Astron. J. \textbf{116} (1998), 1009-1038.

\bibitem{Peebles:2002gy}
P.~J.~E.~Peebles and B.~Ratra,
Rev. Mod. Phys. \textbf{75} (2003), 559-606.

\bibitem{Boehm:2000gq}
C.~Boehm, P.~Fayet and R.~Schaeffer,
Phys. Lett. B \textbf{518} (2001), 8-14.

\bibitem{Guth:1980zm}
A.~H.~Guth,
Phys. Rev. D \textbf{23} (1981), 347-356.

\bibitem{DiValentino:2019qzk}
E.~Di Valentino, A.~Melchiorri and J.~Silk,
Nature Astron. \textbf{4} (2019) no.2, 196-203.

\bibitem{Abdalla:2022yfr}
E.~Abdalla, \textit{et al.}
JHEAp \textbf{34} (2022), 49-211.

\bibitem{Perivolaropoulos:2021jda}
L.~Perivolaropoulos and F.~Skara,
New Astron. Rev. \textbf{95} (2022), 101659.

\bibitem{Capozziello:2011et}
S.~Capozziello and M.~De Laurentis,
Phys. Rept. \textbf{509} (2011), 167-321.

\bibitem{Hinterbichler:2011tt}
K.~Hinterbichler,
Rev. Mod. Phys. \textbf{84} (2012), 671-710.

\bibitem{Maartens:2010ar}
R.~Maartens and K.~Koyama,
Living Rev. Rel. \textbf{13} (2010), 5.

\bibitem{Hehl:1976kj}
F.~W.~Hehl, P.~Von Der Heyde, G.~D.~Kerlick and J.~M.~Nester,
Rev. Mod. Phys. \textbf{48} (1976), 393-416.

\bibitem{Hehl:1994ue}
F.~W.~Hehl, J.~D.~McCrea, E.~W.~Mielke and Y.~Ne'eman,
Phys. Rept. \textbf{258} (1995), 1-171.


\bibitem{DeFelice:2010aj}
A.~De Felice and S.~Tsujikawa,
Living Rev. Rel. \textbf{13} (2010), 3.

\bibitem{Arcos:2004zh}
H.~I.~Arcos and J.~G.~Pereira,
Class. Quant. Grav. \textbf{21} (2004), 5193-5202.

\bibitem{BeltranJimenez:2017tkd}
J.~Beltr\'an Jim\'enez, L.~Heisenberg and T.~Koivisto,
Phys. Rev. D \textbf{98} (2018) no.4, 044048.

\bibitem{Aldrovandi:2013wha}
R.~Aldrovandi and J.~G.~Pereira,
Springer, 2013.

\bibitem{Baez:2012bn}
J.~C.~Baez and D.~K.~Wise,
Commun. Math. Phys. \textbf{333} (2015) no.1, 153-186.

\bibitem{Hohmann:2017duq}
M.~Hohmann, L.~J\"arv, M.~Kr\v{s}\v{s}\'ak and C.~Pfeifer,
Phys. Rev. D \textbf{97} (2018) no.10, 104042.



\bibitem{Ferraro:2006jd}
R.~Ferraro and F.~Fiorini,
Phys. Rev. D \textbf{75} (2007), 084031.

\bibitem{Cai:2015emx}
Y.~F.~Cai, S.~Capozziello, M.~De Laurentis and E.~N.~Saridakis,
Rept. Prog. Phys. \textbf{79} (2016) no.10, 106901.

\bibitem{Nester:1998mp}
J.~M.~Nester and H.~J.~Yo,
Chin. J. Phys. \textbf{37} (1999), 113


\bibitem{BeltranJimenez:2018vdo}
J.~Beltr\'an Jim\'enez, L.~Heisenberg and T.~S.~Koivisto,
JCAP \textbf{08}, 039 (2018)



\bibitem{BeltranJimenez:2017tkd}
J.~Beltr\'an Jim\'enez, L.~Heisenberg and T.~Koivisto,
Phys. Rev. D \textbf{98} (2018) no.4, 044048.

\bibitem{Xu:2019sbp}
Y.~Xu, G.~Li, T.~Harko and S.~D.~Liang,
Eur. Phys. J. C \textbf{79} (2019) no.8, 708

\bibitem{BeltranJimenez:2019tme}
J.~Beltr\'an Jim\'enez, L.~Heisenberg, T.~S.~Koivisto and S.~Pekar,
Phys. Rev. D \textbf{101} (2020) no.10, 103507

\bibitem{Lymperis:2022oyo}
A.~Lymperis,
JCAP \textbf{11} (2022), 018.



\bibitem{Lu:2019hra}
J.~Lu, X.~Zhao and G.~Chee,
Eur. Phys. J. C \textbf{79} (2019) no.6, 530.



\bibitem{Atayde:2021pgb}
L.~Atayde and N.~Frusciante,
Phys. Rev. D \textbf{104} (2021) no.6, 064052.


\bibitem{Bajardi:2020fxh}
F.~Bajardi, D.~Vernieri and S.~Capozziello,
Eur. Phys. J. Plus \textbf{135} (2020) no.11, 912.

\bibitem{Agrawal:2022vdg}
A.~S.~Agrawal, B.~Mishra and P.~K.~Agrawal,
Eur. Phys. J. C \textbf{83} (2023) no.2, 113.

\bibitem{Gadbail:2023loj}
G.~N.~Gadbail, A.~Kolhatkar, S.~Mandal and P.~K.~Sahoo,
Eur. Phys. J. C \textbf{83} (2023) no.7, 595.

\bibitem{Bajardi:2023vcc}
F.~Bajardi and S.~Capozziello,
Eur. Phys. J. C \textbf{83} (2023) no.6, 531.

\bibitem{Dimakis:2021gby}
N.~Dimakis, A.~Paliathanasis and T.~Christodoulakis,
Class. Quant. Grav. \textbf{38} (2021) no.22, 225003.

\bibitem{Lin:2021uqa}
R.~H.~Lin and X.~H.~Zhai,
Phys. Rev. D \textbf{103} (2021) no.12, 124001
[erratum: Phys. Rev. D \textbf{106} (2022) no.6, 069902].

\bibitem{DAmbrosio:2021zpm}
F.~D'Ambrosio, S.~D.~B.~Fell, L.~Heisenberg and S.~Kuhn,
Phys. Rev. D \textbf{105} (2022) no.2, 024042.

\bibitem{Junior:2023qaq}
J.~T.~S.~S.~Junior and M.~E.~Rodrigues,
Eur. Phys. J. C \textbf{83} (2023) no.6, 475.

\bibitem{Mustafa:2021ykn}
G.~Mustafa, Z.~Hassan, P.~H.~R.~S.~Moraes and P.~K.~Sahoo,
Phys. Lett. B \textbf{821} (2021), 136612.

\bibitem{Mustafa:2021bfs}
G.~Mustafa, Z.~Hassan and P.~K.~Sahoo,
Annals Phys. \textbf{437} (2022), 168751.

\bibitem{Mishra:2023bfe}
A.~K.~Mishra, Shweta and U.~K.~Sharma,
Universe \textbf{9} (2023) no.4, 161.

\bibitem{Silva:2022pfd}
J.~E.~G.~Silva, R.~V.~Maluf, G.~J.~Olmo and C.~A.~S.~Almeida,
Phys. Rev. D \textbf{106} (2022) no.2, 024033.

\bibitem{Fu:2021rgu}
Q.~M.~Fu, L.~Zhao and Q.~Y.~Xie,
Eur. Phys. J. C \textbf{81} (2021) no.10, 890.

\bibitem{Belchior:2023xgn}
F.~M.~Belchior, A.~R.~P.~Moreira, R.~V.~Maluf and C.~A.~S.~Almeida,
Phys. Lett. B \textbf{843} (2023), 138029.

\bibitem{Heisenberg:2023lru}
L.~Heisenberg,
Phys. Rept. \textbf{1066} (2024), 1-78.

\bibitem{Randall:1999ee}
L.~Randall and R.~Sundrum,
Phys. Rev. Lett. \textbf{83} (1999), 3370.

\bibitem{Randall:1999vf}
L.~Randall and R.~Sundrum,
Phys. Rev. Lett. \textbf{83} (1999), 4690.

\bibitem{DeWolfe:1999cp}
O.~DeWolfe, D.~Z.~Freedman, S.~S.~Gubser and A.~Karch,
Phys. Rev. D \textbf{62} (2000), 046008.

\bibitem{Shiromizu:1999wj}
T.~Shiromizu, K.~i.~Maeda and M.~Sasaki,
Phys. Rev. D \textbf{62} (2000), 024012.

\bibitem{Binetruy:1999ut}
P.~Binetruy, C.~Deffayet and D.~Langlois,
Nucl. Phys. B \textbf{565} (2000), 269.

\bibitem{Israel:1966rt}
W.~Israel,
Nuovo Cim. B \textbf{44S10} (1966), 1
[erratum: Nuovo Cim. B \textbf{48} (1967), 463].















\bibitem{Ahmed:2013lea}
A.~Ahmed, B.~Grzadkowski and J.~Wudka.
JHEP \textbf{04} (2014), 061.


\bibitem{Hendi:2020qkk}
S.~H.~Hendi, N.~Riazi and S.~N.~Sajadi,
Phys. Rev. D \textbf{102} (2020) no.12, 124034

\bibitem{Bernardini:2014vba}
A.~E.~Bernardini, R.~T.~Cavalcanti and R.~da Rocha,
Gen. Rel. Grav. \textbf{47} (2015) no.1, 1840.

\bibitem{daSilva:2016ntp}
P.~M.~L.~T.~da Silva, A.~de Souza Dutra and J.~M.~Hoff da Silva,
Phys. Lett. B \textbf{774} (2017), 482-488.

\bibitem{Gremm:1999pj}
M.~Gremm,
Phys. Lett. B \textbf{478} (2000), 434.

\bibitem{Afonso:2006gi}
V.~I.~Afonso, D.~Bazeia and L.~Losano,
Phys. Lett. B \textbf{634} (2006), 526.

\bibitem{Giovannini:2007xb}
M.~Giovannini,
Phys. Rev. D \textbf{76} (2007), 124017.

\bibitem{Kadosh:2012ru}
A.~Kadosh, A.~Davidson and E.~Pallante,
Phys. Rev. D \textbf{86} (2012), 124015.

\bibitem{Guha:2018hbr}
S.~Guha and P.~Bhattacharya,
Grav. Cosmol. \textbf{24} (2018) no.3, 274.

\bibitem{Carroll:2004st}
S.~M.~Carroll,
Cambridge University Press, 2019.











\end{thebibliography}
\end{document}